Investigation on the reported superconductivity in intercalated black phosphorus

Hanming Yuan[a], Liangzi Deng[a], Bing Lv[b], Zheng Wu[a], Ze Yang[a], Sheng Li[b], Shuyuan Huyan[a], Yizhou Ni[a], Jingying Sun[a], Fei Tian[a], Dezhi Wang[a], Hui Wang[a], Shuo Chen[a], Zhifeng Ren[a], and Ching-Wu Chu[a]*

[a]Department of Physics and Texas Center for Superconductivity, University of Houston, Houston, TX 77204-5005, USA

[b]Department of Physics, University of Texas at Dallas, Richardson, TX 75080

**Abstract**

Superconductivity intrinsic to the intercalated black phosphorus (BP) with a transition temperature $T_c$ of 3.8 K, independent of the intercalant, whether an alkali or an alkaline earth element, has been reported recently by R. Zhang et al. (2017). However, the reported $T_c$ and the field effect on the superconducting (SC) transition both bear great similarities to those for the pure Sn, which is commonly used for BP synthesis under the vapor transport method. We have therefore decided to determine whether a minute amount of Sn is present in the starting high purity BP crystals and whether it is the culprit for the small SC signal detected. Energy-dispersive X-ray spectroscopy results confirmed the existence of Sn in the starting high purity BP crystals purchased from the same company as in R. Zhang et al. (2017). We have reproduced the SC transition at 3.8 K in Li- and Na-intercalated BP crystals that contain minute amounts of Sn when prepared by the vapor transport method but not in BP crystals that are free of Sn when prepared by the high-pressure method. We have therefore concluded that the SC transition reported by R. Zhang et al. (2017) is associated with the Sn but not intrinsic to the intercalated BP crystals.

## 1. Introduction

Building on the work on graphene and a few other layered materials that followed, the two-dimensional layered black phosphorus (BP) has attracted great interest recently because of its scientific significance and device potential. Of particular interest is the tunability of its physical properties through varying the band structures by strain, electric field, number of layers, and intercalation. Superconductivity has also been predicted in electron-doped monolayer [1] and Li-intercalated bilayer phosphorene [2]. This culminated in the recent report by Zhang *et al.* [3] of superconductivity intrinsic to the intercalated BP with a transition temperature $T_c$ of 3.8 K, independent of the intercalant, whether an alkali or an alkaline earth element. Indeed, this is an observation of a highly unusual superconducting state that is independent of the valence, the content, and the size of the dopant, in contrast to previous understanding of known superconductors. The authors attribute this so-called universal superconductivity to the heavily doped phosphorene layers with the intercalated layers serving as charge reservoirs, similar to the modulation doping in layered high temperature superconductors. The significance of the report is self-evident, if proven.

We have therefore examined the reported results [3] carefully and systematically. The intercalant-independent $T_c$ of BP and the apparent small superconducting volume fraction of all samples investigated led us to the obvious question, *i.e.*, could the observed superconductivity be caused by a small superconducting contamination in the samples? Given the reputation of the group, contamination introduced during sample preparation seems to be highly unlikely. However, it is rather intriguing to find that the reported $T_c$ of 3.8 K and the field effect on the superconducting transition both appear to be similar to those for the pure Sn [4]. The magnetic anisotropy reported is also too small to be consistent with the model proposed. It is also known that most of the commercially available BP crystals are prepared by transforming the red phosphorus through the chemical vapor transport technique with transport agents consisting of Sn or Sn-related compounds [5-8]. We have therefore decided to determine whether a minute amount of Sn is present in the starting BP crystals and, if yes, whether it is the culprit for the small superconducting signal detected.

## 2. Experimental

To determine if Sn-contamination can be introduced from chemical vapor transport synthesis, we have therefore started with three different BP crystal sources (BP-1, BP-2, and BP-3): BP-1 was purchased from Smart Elements in Germany, from which Zhang *et al.* obtained their BP crystals of the same purity (99.998%); BP-2 was prepared in our lab by the standard vapor transport method at ambient pressure; and BP-3 was made in our lab by the well-known high pressure technique [9].

2.1 Material synthesis

BP-2 crystals were prepared by converting the red phosphorus to black phosphorus by sealing appropriate amounts of red phosphorus (99.999%, Aldrich, 350 mg), Sn (99.99+%, Alfa Aesar, 35 mg), and $I_2$ (99.8+%, Fisher Scientific, 30 mg) in an evacuated quartz tube (12.7 mm in diameter, 127 mm in length), which was then placed in a tubular furnace with a temperature of 620 °C at the center and 27 °C at the end. Crystals of sizes up to ~ 3 mm × 1 mm × 0.5 mm were harvested at the cold end of the quartz tube.

BP-3 crystals were achieved by transforming red phosphorus powder (99.99+%, Aldrich, 0.28 g) wrapped in Au-foil under 1.5 GPa at 750 °C for 30 minutes followed by quenching to room temperature.

2.2 Intercalation

Two different methods were employed to intercalate these BP crystals: A) n-Butyllithium treatment for Li-intercalant, and B) electrochemical treatment for Li- and Na-intercalants. According to method A, the BP crystals were immersed for one week in the 1.33 M n-butyllithium (Alfa Aesar) in hexane (Aldrich) solution in a beaker covered by parafilm in an Ar atmosphere at room temperature. According to method B, a galvanostatic cell with a BP-composite electrode as a cathode and a metal anode immersed in an electrolyte was discharged to 0.01 V at a current of 2 µA. For Li-intercalation the anode was Li-metal and the electrolyte was 1 M $LiPF_6$ in a 1:1 diethyl carbonate/ethylene carbonate mixed solution with Celgard polypropylene as the separator. For Na-intercalation, the anode was Na-metal and the electrolyte was 1 M $NaClO_4$ in a 1:1 propylene carbonate/ethylene carbonate mixed solution with glass fiber as the separator.

## 3. Results and discussion

X-ray diffraction (XRD) results with distinct (0$\ell$0)- peaks characteristic of single crystals of the starting BP-1, -2, and -3 (with weak peaks from other orientations and a trace amount of $Au_2P_3$ from the wrapping during synthesis for BP-3), all with a b ~ 10.472 Å, in agreement with published data, are shown in Fig. 1. Energy-dispersive X-ray spectroscopy (EDS) has also been carried out on all these starting BP crystals, as shown in Fig. 2. The results show traces of Sn and I in BP-1 and -2, but not in BP-3, as expected. Detailed magnetic measurements ($\chi_\perp$s) with the measuring field perpendicular to the ac-plane to maximize the signal did not display any sign of superconductivity within the MPMS3 sensitivity limit of $10^{-8}$ emu down to 2 K in these starting BP crystals.

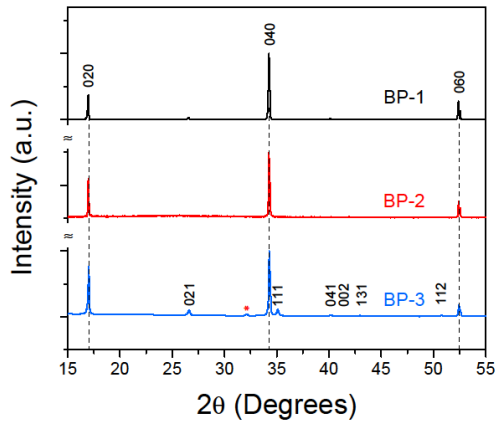

**Fig. 1.** X-ray diffraction spectra of starting BP-1, -2, and -3. * indicates the known peak for $Au_2P_3$.

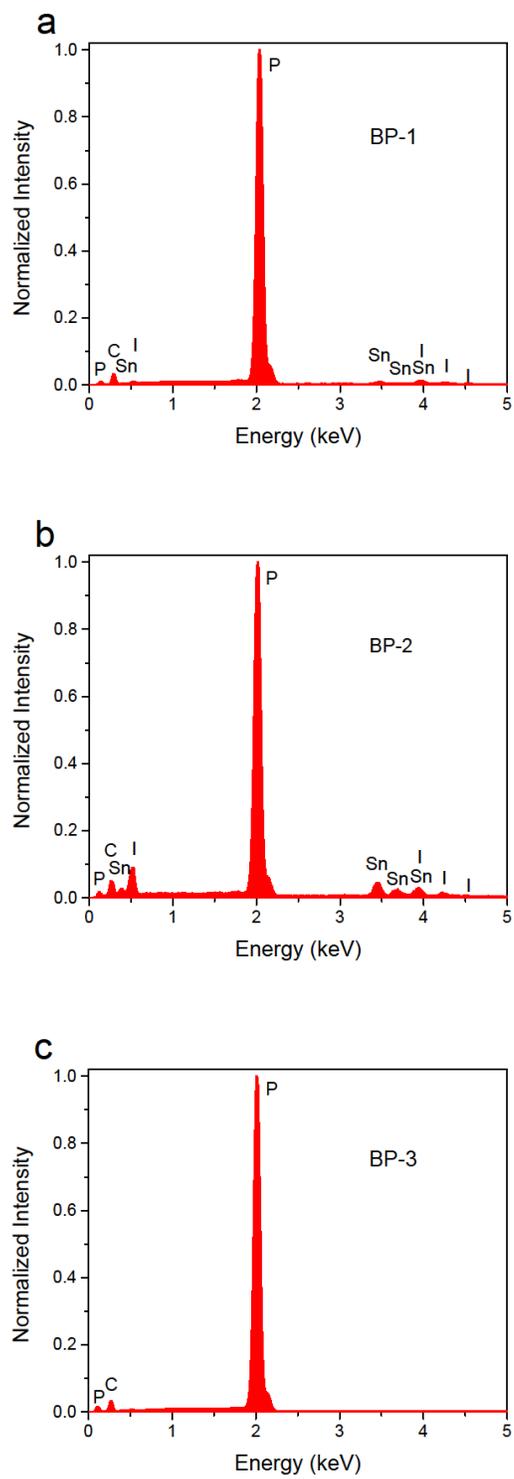

**Fig. 2.** EDS analysis for starting BP crystals. (a) BP-1. (b) BP-2. (c) BP-3.

Intercalations were then performed on these starting BP crystals as described in section 2.2. XRD of the Li- and Na-intercalated BP samples all show the same diffraction patterns within our resolution as those before intercalation, as shown in Fig. 3. The exposure to air does not seem to affect the XRD patterns, in contrast to that previously reported [3], probably due to slight misalignment of the crystals in the previous report.

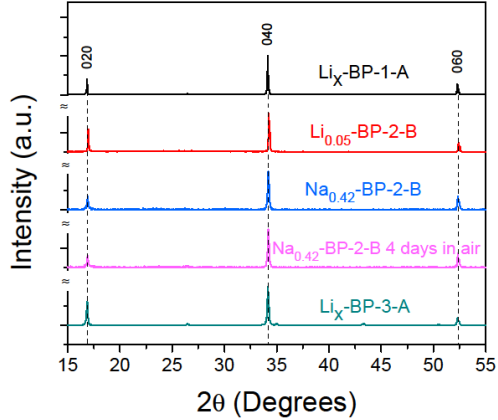

**Fig. 3.** X-ray diffraction spectra of Li- or Na-intercalated BP-1, -2, and -3. The same Na-intercalated BP-2 was measured before and after exposure to air for 4 days.

The magnetic measurements, $\chi_\perp$s, with the measuring field perpendicular to the ac-plane, show a small but distinct and sharp superconducting transition with a $T_c \sim 3.8$ K of up to $\sim 1\%$ volume fraction in the Li- and Na-intercalated BP-1 and -2 but not in the BP-3, as shown in Figs. 4a and b, consistent with the absence of the resistive sign of superconductivity shown in Fig. 5. In Fig. 4c, an externally applied magnetic field is shown to suppress the superconducting transition progressively to below 2 K above ~300 Oe, suggesting a critical field of ~ 300 Oe. Similar to Zhang *et al*., we found that the $T_c$ is ~ 3.8 K in BP-1 and BP-2 intercalated with Li by method A or B, and with Na by method B, independent of the intercalant or the amount of intercalant. For comparison, the superconducting transition of Sn is also included in Figs. 4a and b. The similarity between the intercalated BP-1 and -2 and Sn is evident. Field effects on the superconducting transitions of Li-intercalated BP-2-B and Sn are depicted in Fig. 4c. The similarity between the two again is clear. The magnetic anisotropy is small ~ 1.3, as shown in Fig 6, in contrast to that expected of the model proposed [3].

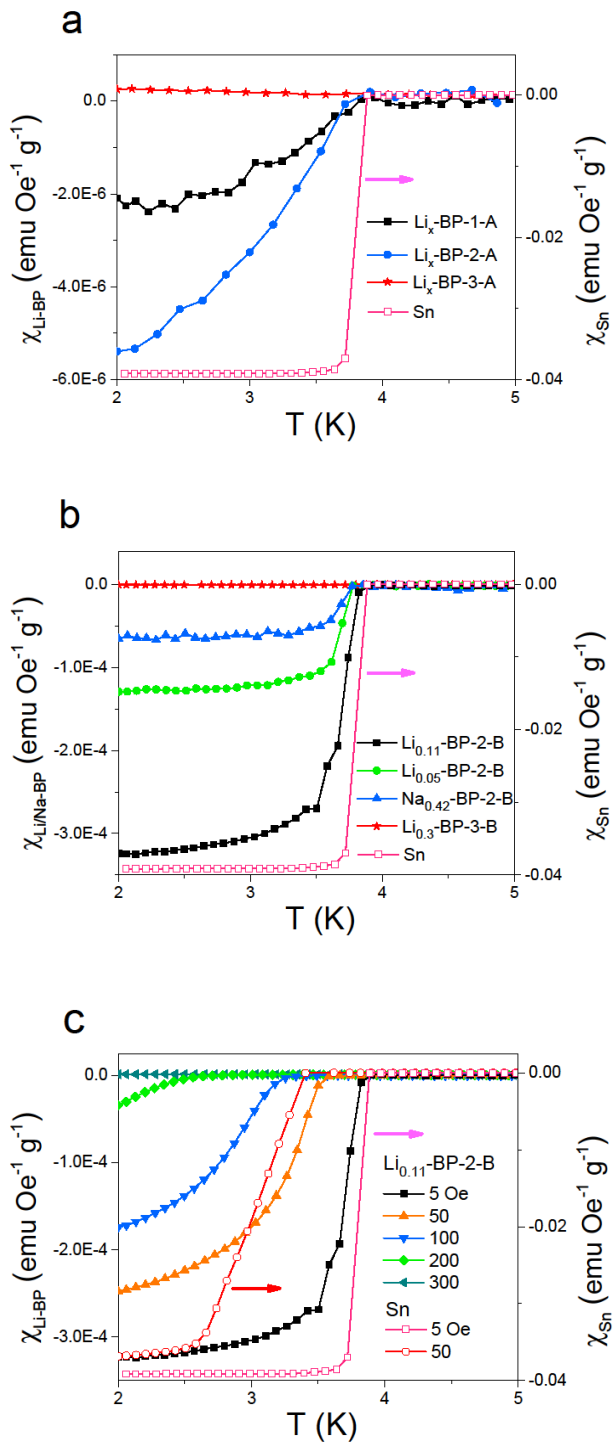

**Fig. 4.** Comparison between intercalated BP and Sn. Temperature dependence of $\chi_\perp$s for (a) n-Butyllithium-intercalated BP-1, -2, and -3, and Sn, and (b) electrochemically intercalated BP-2 and -3 and Sn. (c) Field effects on Li-intercalated BP-2 and Sn. All data shown are ZFC $\chi_\perp$s.

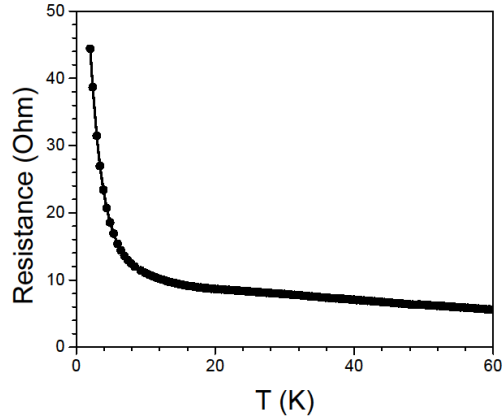

**Fig. 5.** Resistance vs. temperature of Li-intercalated BP-2.

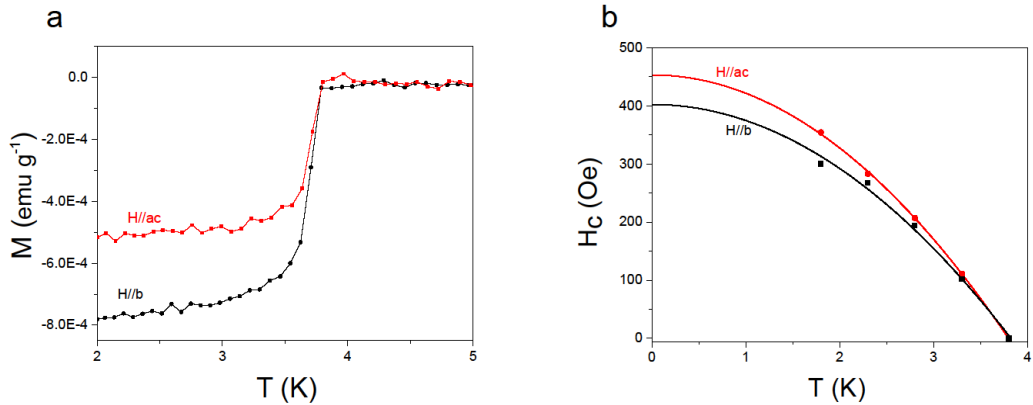

**Fig. 6.** Magnetic anisotropy of Li-intercalated BP-2. (a) Parallel and perpendicular magnetization as a function of temperature at 5 Oe. ZFC curves are shown. (b) Parallel and perpendicular critical fields as a function of temperature.

## 4. Conclusion

In conclusion, we have reproduced the superconducting transition at 3.8 K in Li- and Na-intercalated BP crystals that contain minute amounts of free Sn (reduced from the residual non-superconducting Sn compounds by Li or Na) when prepared by the vapor transport technique but not in BP crystals that are free of Sn when prepared by the high pressure technique. The superconducting transition takes place at the same temperature as pure Sn and the field effect on the transition is similar to that on Sn. The magnetic anisotropy of the superconducting state in the Li- and Na-intercalated BP is rather small. We have therefore concluded that the superconducting transition reported by Zhang *et al.* is associated with the Sn but not intrinsic to the intercalated BP.

**Acknowledgement**

The work in Houston is supported in part by U.S. Air Force Office of Scientific Research Grant No. FA9550-15-1-0236, the T.L.L. Temple Foundation, the John J. and Rebecca Moores Endowment, and the State of Texas through the Texas Center for Superconductivity at the University of Houston.